\documentclass[10pt]{iopart}

\begin{document}

\letter{Specific heat of heavy fermion CePd$_2$Si$_2$ in high magnetic fields}

\author{I Sheikin\dag, Y Wang\dag, F Bouquet\dag, P Lejay\ddag\ and A Junod\dag
}

\address{\dag\ University of Geneva, DPMC, 24 Quai Ernest-Ansermet, 1211 Geneva 4, Switzerland}

\address{\ddag\ CRTBT, CNRS, BP 166, 38042 Grenoble, France}


\begin{abstract}

We report specific heat measurements on the heavy fermion compound CePd$_2$Si$_2$ in magnetic fields up to 16$\:$T and in the temperature
range 1.4---16$\:$K. A sharp peak in the specific heat signals the antiferromagnetic transition at $T_N\sim9.3\:$K in zero field. The
transition is found to shift to lower temperatures when a magnetic field is applied along the crystallographic $a$-axis, while a field
applied parallel to the tetragonal $c$-axis does not affect the transition. The magnetic contribution to the specific heat below $T_ N$ is
well described by a sum of a linear electronic term and an antiferromagnetic spin wave contribution. Just below $T_N$, an additional
positive curvature, especially at high fields, arises most probably due to thermal fluctuations. The field dependence of the coefficient of
the low temperature linear term, $\gamma_0$, extracted from the fits shows a maximum at about 6$\:$T, at the point where an anomaly was
detected in susceptibility measurements. The relative field dependence of both $T_N$ and the magnetic entropy at $T_N$ scales as
$[1-(B/B_0)^2]$ for $B\parallel a$, suggesting the disappearance of antiferromagnetism at $B_0\sim42\:$T. The expected suppression of the
antiferromagnetic transition temperature to zero makes the existence of a magnetic quantum critical point possible.

\end{abstract}





The heavy fermion system CePd$_2$Si$_2$ has recently become a subject of considerable experimental and theoretical interest. This compound
undergoes an antiferromagnetic transition at $T_N\sim9\:$K with a static moment of 0.6~$\mu_B$ at low temperature. Its spin configuration
consists of ferromagnetic (110) planes with spins normal to the planes and alternating in directions along the spin axis.

The existence of a quantum critical point at a pressure of about 30$\:$kbar, where the antiferromagnetic transition temperature is
suppressed to zero, was established a long time ago \cite{Thompson}. It is, however, only recently that Grosche {\it et al} \cite{Grosche}
discovered the appearance of superconductivity in a small pressure range in the vicinity of the quantum critical point. This pioneering
work gave rise to a new wave of interest in this compound. Recently, several experimental studies of CePd$_2$Si$_2$ were performed both at
ambient and high pressure on different samples and using different experimental techniques \cite{Raymond,Sheikin,Demuer1,Demuer2}. These
works confirmed the emergence of superconductivity close to the critical pressure and gave an insight into the nature of the
superconducting ground state. It has been also shown that non-Fermi liquid behavior at low temperature appears close to the critical
pressure. Such behavior observed in resistivity, magnetization and specific heat measurements is a common trend in heavy fermion compounds
in the vicinity of a magnetic instability induced by application of hydrostatic pressure.

A different way to tune a heavy fermion compound to a quantum critical point is by applying a high magnetic field.
Field suppression of antiferromagnetism, like pressure, does not induce any disorder and, like pressure, opens up a new
dimension in the phase diagram for study. Such a suppression of the magnetically ordered ground state by high magnetic
field was found in CeCu$_{5.2}$Ag$_{0.8}$ \cite{Heuser}, YbCu$_{5-x}$Al$_x$ \cite{Bauer,Seuring}, YbRh$_2$Si$_2$
\cite{Custers} and CeCu$_{5.2}$Au$_{0.8}$ \cite{Löhneysen}. In all these cases, unusual non-Fermi liquid behavior was
also found near the magnetic quantum critical point. There are, however, other systems, e.g. CePtSi$_{0.4}$Ge$_{0.6}$
\cite{Heuser}, that show Fermi-liquid behavior upon field suppression of $T_N$ to zero.

Two experimental investigations of CePd$_2$Si$_2$ at high magnetic field were recently reported. The first publication \cite{Abe} reports
magnetization measurements up to 28$\:$T at 4.2$\:$K with the magnetic field applied along both $a$ and $c$ crystallographic axes. No
anomalies have been seen for either direction of the magnetic field. The other work \cite{Heuser} presents specific heat measurements at 0,
13 and 28.9$\:$T on a polycrystalline sample. At 28.9$\:$T, the antiferromagnetic transition was found to be broadened with both the peak
in the specific heat and the transition midpoint suppressed to lower temperatures though the onset did not shift. This left the situation
in this compound somewhat obscure, and no further high field investigations have been reported.

We have, therefore, decided to reexamine the experimental situation in this material. We have performed specific heat measurements on a
single crystal of CePd$_2$Si$_2$ in magnetic fields up to 16$\:$T in the temperature range 1.4---16$\:$K.


The specific heat measurements were performed on a single crystal of CePd$_2$Si$_2$, grown in a triarc furnace by the
Czochralski method. Further details of the sample preparation are given elsewhere \cite{vanDijk1}. The sample, with a
mass of 15$\:$mg, was mounted on a small sapphire plate. A carbon film with four contacts made of phosphor bronze was
painted on the other side of the plate. The whole set-up including two thermometers and a heater was sealed inside a
vacuum sample chamber. A carbon glass resistive thermometer was used to regulate the temperature inside the sample
chamber. The temperature was measured by a Cernox thermometer calibrated in field. A thermal relaxation technique was
used to measure the specific heat \cite{Relaxation}. To extract the specific heat of the sample, the contribution of
the sapphire, electrical wires and grease used as an adhesive was measured and subtracted. Magnetic fields up 16$\:$T
were generated by a superconducting coil. The temperature was varied from 1.4 to 16$\:$K using a standard variable
temperature insert (VTI).


Figure~\ref{zero pressure} shows the zero-field specific heat. The data are in good agreement with previous measurements
\cite{vanDijk1,Steeman}. The antiferromagnetic transition manifests itself by a sharp, almost step-like increase of the specific heat at
9.3$\:$K. The magnetic contribution, $C_{\rm mag}$, to the specific heat is obtained by subtracting the data of the non-magnetic reference
system LaPd$_2$Si$_2$ \cite{Besnus} (also shown in figure~\ref{zero pressure}). At low temperature, LaPd$_2$Si$_2$ shows the classical
behavior $C=\gamma_r T+\beta_r T^3$ with $\gamma_r=6.0\:$mJ/molK$^2$ and $\beta_r=0.267\:$mJ/molK$^4$. This is to be compared with
$C/T=160\:$mJ/molK$^2$ found for CePd$_2$Si$_2$ at 1.4$\:$K, the lowest temperature of our measurements (see, however, the discussion below
for fits of $C/T$ as $T\rightarrow0$).

Above the antiferromagnetic transition, $C_{\rm mag}/T$ increases with temperature decrease due to the Kondo spin fluctuations. Below
$T_N$, $C_{\rm mag}(T)$ can be described in terms of spin fluctuations and an additional contribution from the antiferromagnetic spin waves
\cite{vanDijk1}. We have integrated $C_{\rm mag}/T$ to obtain the magnetic entropy. Since the lowest temperature of our measurements was
1.4$\:$K, we extrapolated the data below 1.4$\:$K using the empirical temperature dependence $C=\gamma T+\beta T^3$ discussed below. The
calculated magnetic entropy is shown in the inset of figure~\ref{zero pressure}. The entropy at the transition temperature reaches about
75\% of $R\ln2$, the value expected for a doublet ground state \cite{vanDijk1}. This is not necessarily surprising and is likely to imply
that the ordered moment in the antiferromagnetic phase is somewhat compensated by the Kondo interactions. At 22~K, the highest temperature
of the zero-field measurement, the magnetic entropy reaches 95\% of $R\ln2$.

When a magnetic field is applied parallel to the crystallographic $a$-axis, the transition gradually moves to lower
temperatures remaining, however, as sharp as in zero field as shown in figure~\ref{under-pressure}. Conversely, a
magnetic field of 14$\:$T applied along the tetragonal $c$-axis was not found to affect the transition temperature (see
the inset of figure~\ref{under-pressure}). This implies the existence of strong magnetic anisotropy between the basal
plane and the tetragonal $c$-axis of the compound. Such anisotropy is consistent with the magnetic structure of the
material with its spins aligned in the basal plane. This anisotropy explains the result of a previous report obtained
on a polycrystal \cite{Heuser}, where at 29$\:$T, the transition was found to be much broader with the peak in $C/T$
suppressed to about 5$\:$K but without any shift of the onset. Indeed, in the case of a polycrystal, the part of the
sample more or less aligned with the field along the $c$-axis shows no change and that is why the onset of the
transition does not move. On the other hand, the other field directions suppress the transition which leads both to a
shift of the specific heat peak and the transition broadening.

We have tried to fit the data below $T_N$ by the commonly used empirical fitting functions $C_{\rm mag}=\gamma T+A\exp(-\Delta /T)$ for a
spin gap \cite{Fisher} and $C_{\rm mag}=\gamma T+\beta T^3$ for antiferromagnetic spin waves \cite{vanDijk1}. While both of them fit well
the low temperature part of the data, neither of them succeeded in providing a satisfactory fit over the whole temperature range. We found,
however, that the magnetic contribution to the specific heat is well described by a model composed of a linear electronic contribution as
above, $\gamma_0T$, plus a term that accounts for the contribution from antiferromagnetic spin waves with the dispersion relation
$\omega=\sqrt{\Delta^2+Dk^2}$ \cite{deMedeiros}:
\begin{equation}
C_{\rm mag}=\gamma_0T+\alpha\Delta^{7/2}T^{1/2}\rme^{-\Delta /T}[1+(39/20)(T/\Delta)+(51/32)(T/\Delta)^2]
\end{equation}
Here $\Delta$ is the spin-wave gap, and $\alpha$ is related to the spin-wave stiffness $D$ by $\alpha \propto D^{-3/2}$. As shown in
figure~\ref{fits} for $B=0$ and 16$\:$T, the above equation yields a good fit to the data, except for a small temperature range just below
$T_N$, which was ignored when fitting the data. The failure of the fit over this small temperature range is due to the existence of a
positive curvature just below $T_N$. This curvature becomes gradually stronger with field as can be seen in figure~\ref{fits}. We associate
this curvature with thermal fluctuations that are expected to play a role in the vicinity of the transition, and to become stronger in
magnetic field.

The field dependence of the low temperature electronic contribution to the specific heat extracted from the fits is shown in
figure~\ref{gamma}. As one can see, the low temperature Sommerfield coefficient, $\gamma_0$, passes through a maximum at about 6$\:$T
before starting to decrease monotonically. This maximum matches a clear anomaly observed in susceptibility measurements when the magnetic
field was also applied along the crystallographic $a$-axis \cite{Semeno}. The anomaly can be associated with a spin-flop process. Indeed,
the behavior of $\gamma_0$ in magnetic field we find here is very similar to that of the coefficient $A$ of the $T^2$ term of the
resistivity found by McDonough and Julian in CePb$_3$ around a spin-flop transition \cite{McDonough&Julian}. Note that the coefficient $A$
also reflects the many-body enhancement, the ratio $A/\gamma_0$ having an universal value in heavy fermion systems, as pointed out by
Kadowaki and Woods \cite{Kadowaki&Woods}.

The field-dependent values of $T_N$ have been determined by an equal-entropy construction with an ideal step
transition. Figure~\ref{T vs P} shows the field dependence of the transition temperature and the magnetic entropy at
$T_N$. Both follow a simple scaling relation $[1-(B/B_0)^2]$. Here $B_0$ corresponds to the field necessary to suppress
the antiferromagnetic order, and is found to be $(41.5\pm0.6)\:$T. This scaling relation explains quantitatively why no
anomalies were observed in the magnetization measurements up to 28$\:$T at $T=4.2\:$K \cite{Abe}. According to the
above formula, at $T=4.2\:$K the transition should occur at about 31$\:$T for $B\parallel a$, while in \cite{Abe} the
highest applied field was only 28$\:$T. The same scaling behavior was found for another heavy fermion system
URu$_2$Si$_2$ \cite{vanDijk2}, where the value of $B_0$ was found to be in relatively good agreement with the results
obtained from resistivity, thermal expansion and magnetization measurements.


In conclusion, we have shown that the antiferromagnetic transition in CePd$_2$Si$_2$ shifts to lower temperatures in magnetic fields
applied in the basal plane. The absence of influence of the magnetic field parallel to the tetragonal $c$-axis on the transition
temperature points to the existence of a strong magnetic anisotropy, which is a common trend for heavy fermion compounds. Analysis of the
field dependence of $T_N$ and the magnetic entropy suggests the suppression of the antiferromagnetic ground state at about 42$\:$T, rising
the possibility of the existence of a quantum critical point at this field. This might make this compound a new member of a growing family
of heavy fermion antiferromagnets which can be tuned by a magnetic field through the magnetic quantum critical point, where the N\'eel
temperature vanishes. This prediction calls for further experiments at higher, probably pulsed fields. The magnetic specific heat below
$T_N$ is best described by the sum of an electronic contribution and antiferromagnetic spin waves. The fit, however, breaks down just below
$T_N$ where thermal fluctuations give rise to a positive curvature. The low temperature coefficient of the electronic term of the specific
heat extracted from the fits is found to have a maximum at about 6$\:$T. This suggests the existence of a magnetic transition, presumably a
spin-flop, the idea being also supported by the results of the magnetization measurements \cite{Semeno}.

\ack{ We thank Dr. A~Semeno for communication of magnetization measurement results prior to publication. This work was supported by Fonds
National Suisse de la Recherche Scientifique. }

\Figures

\begin{figure}
\caption{\label{zero pressure}Specific heat $C/T$ of CePd$_2$Si$_2$ in zero magnetic field. The anomaly at $T_N=9.3\:$K signals the
antiferromagnetic transition. For comparison, the specific heat of the non-magnetic reference system LaPd$_2$Si$_2$ is shown \cite{Besnus}.
The inset shows the magnetic entropy, $S_{\rm mag}/R$, also in zero field.}
\end{figure}

\begin{figure}
\caption{\label{under-pressure}Specific heat $C$ of CePd$_2$Si$_2$ around the antiferromagnetic phase transition in magnetic fields of 0,
4, 8, 12 and 16$\:$T ($B\parallel a$). Both the transition temperature, $T_N$, and the jump in the specific heat decrease with magnetic
field. The inset shows the transition in zero field (closed circles) and at 14$\:$T applied along the $c$-axis (open circles). The two
curves are practically indistinguishable as the transition  does not shift in fields parallel to the $c$-axis.}
\end{figure}

\begin{figure}
\caption{\label{fits}Magnetic specific heat $C_{\rm mag}/T$ of CePd$_2$Si$_2$ in zero field and in a magnetic field of 16$\:$T. The solid
curves correspond to a fit taking into account both the electronic contribution and antiferromagnetic spin waves (equation~1). Parameters
obtained from the fit are: $\gamma_0=(0.131\pm 0.003)\:$J/molK$^2$, $\alpha=(1.7\pm 0.1)\times 10^{-3}\:$J/molK$^4$, $\Delta=(4.6\pm
0.2)\:$K for $B=0$ and $\gamma_0=(0.110\pm 0.006)\:$J/molK$^2$,  $\alpha=(2.5\pm 0.3)\times 10^{-3}\:$J/molK$^4$, $\Delta=(4.2\pm 0.3)\:$K
for $B=16\:$T.}
\end{figure}

\begin{figure}
\caption{\label{gamma}Magnetic field dependence of the coefficient of the electronic linear term of the specific heat extracted from the
fits by equation~1. The weak maximum suggests the possibility of a magnetic phase transition, presumably a spin-flop. The line is a guide
for the eyes. Errors bars correspond to the standard errors obtained from the fit.}
\end{figure}

\begin{figure}
\caption{\label{T vs P}Field dependence of the transition temperature $T_N$ (solid circles, left axis) and the magnetic entropy at $T_{\rm
N}$ (open squares, right axis). The curve represents a scaling function of the form $T_{N0}[1-(B/B_0)^2]$ with $B_0=41.5\:T$.}
\end{figure}

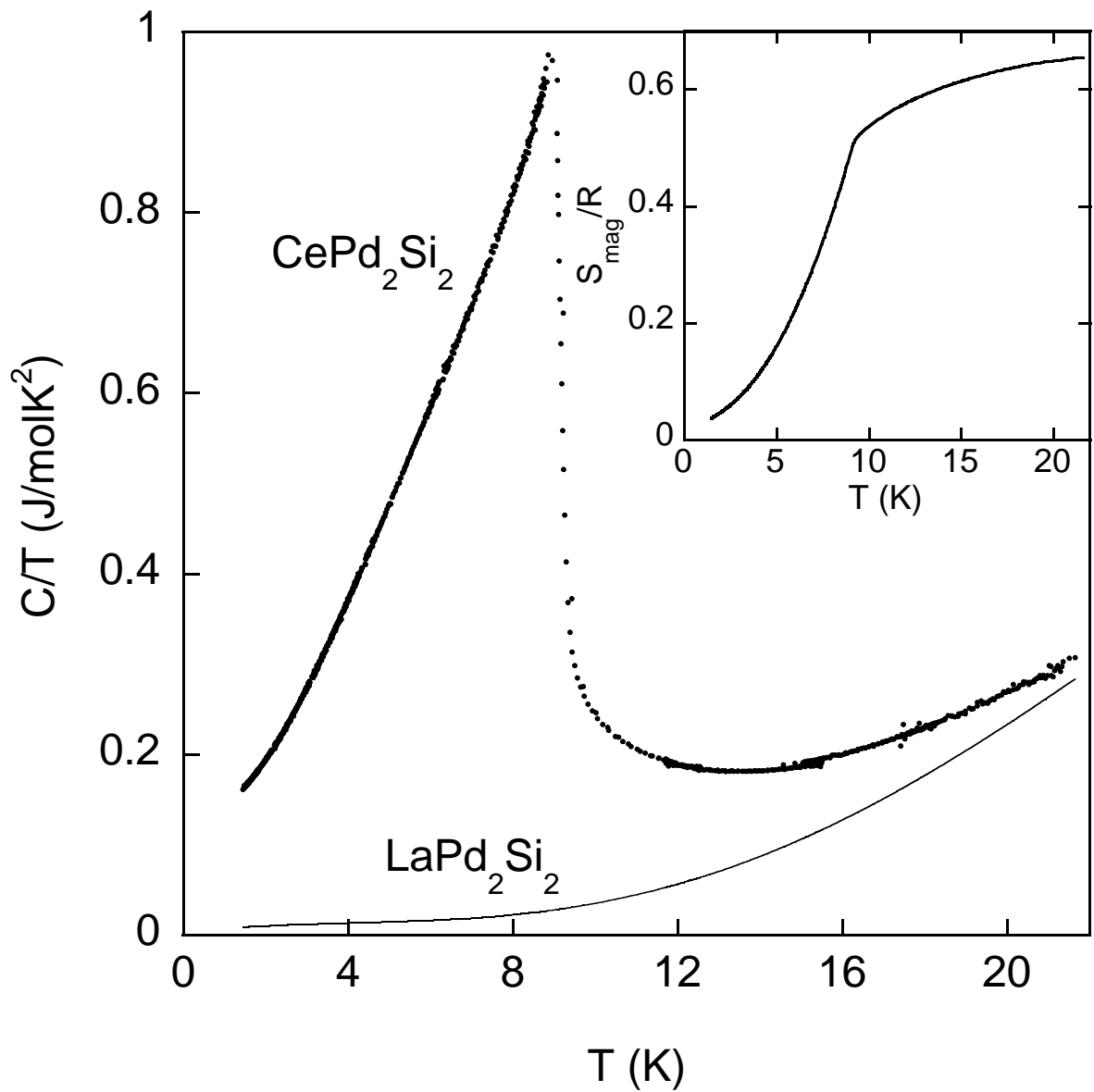
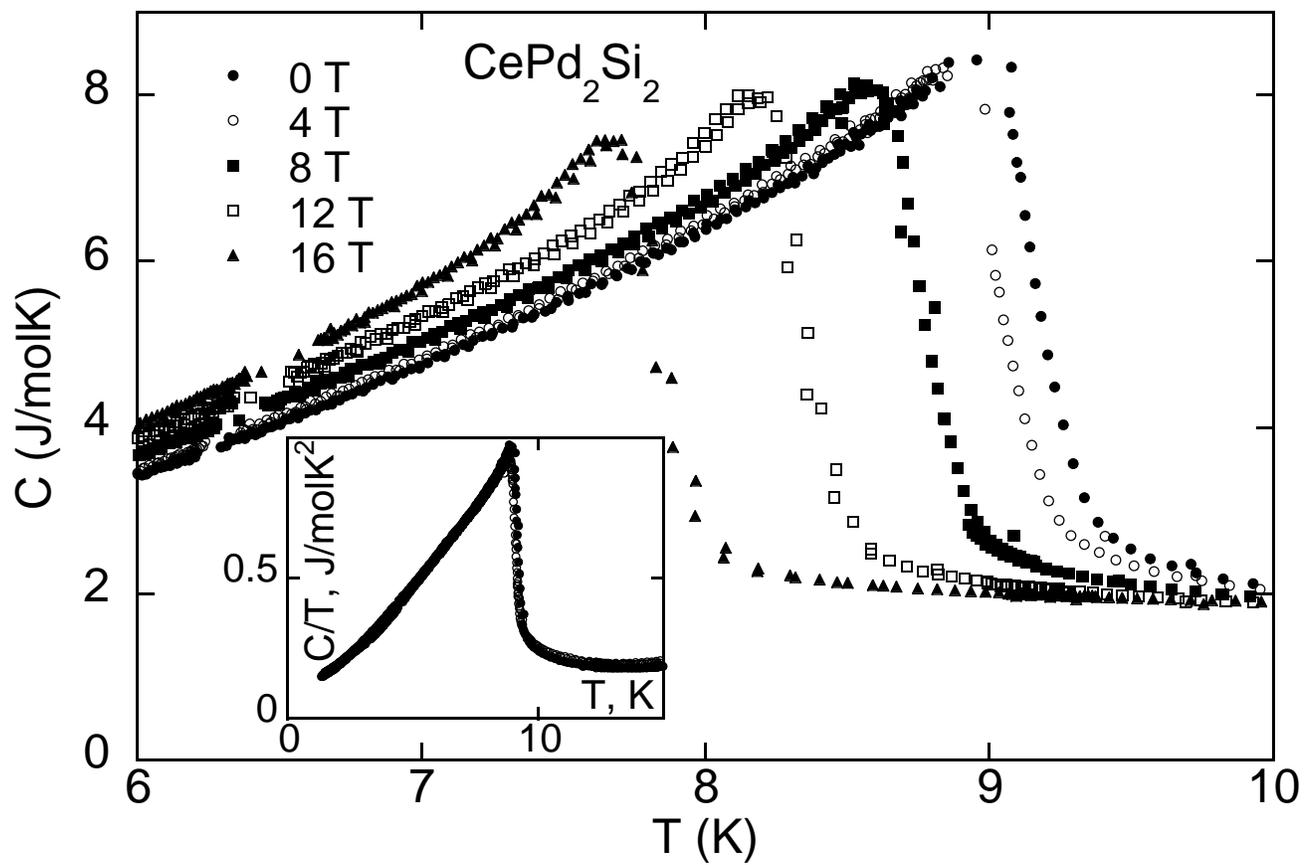
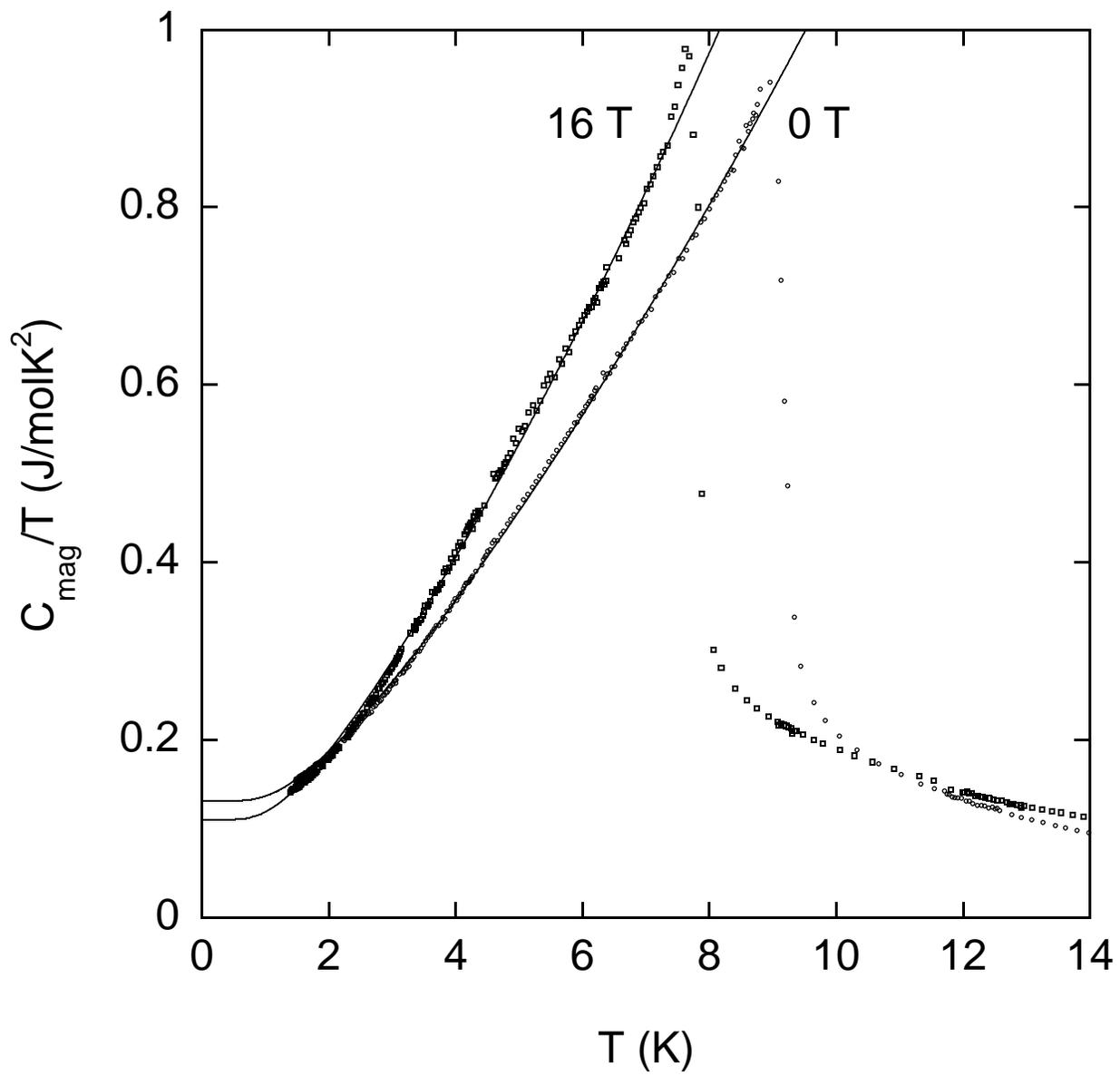
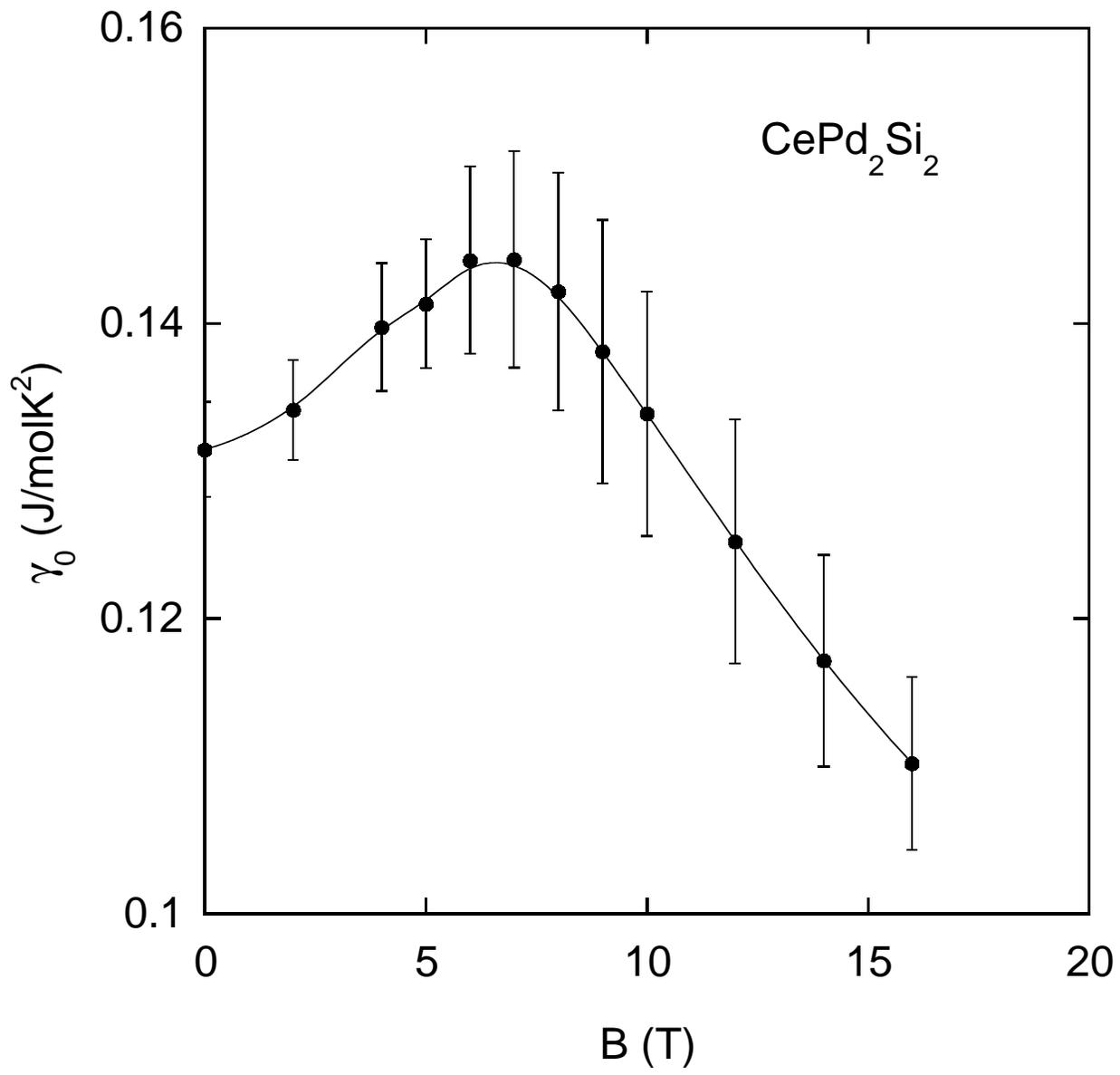
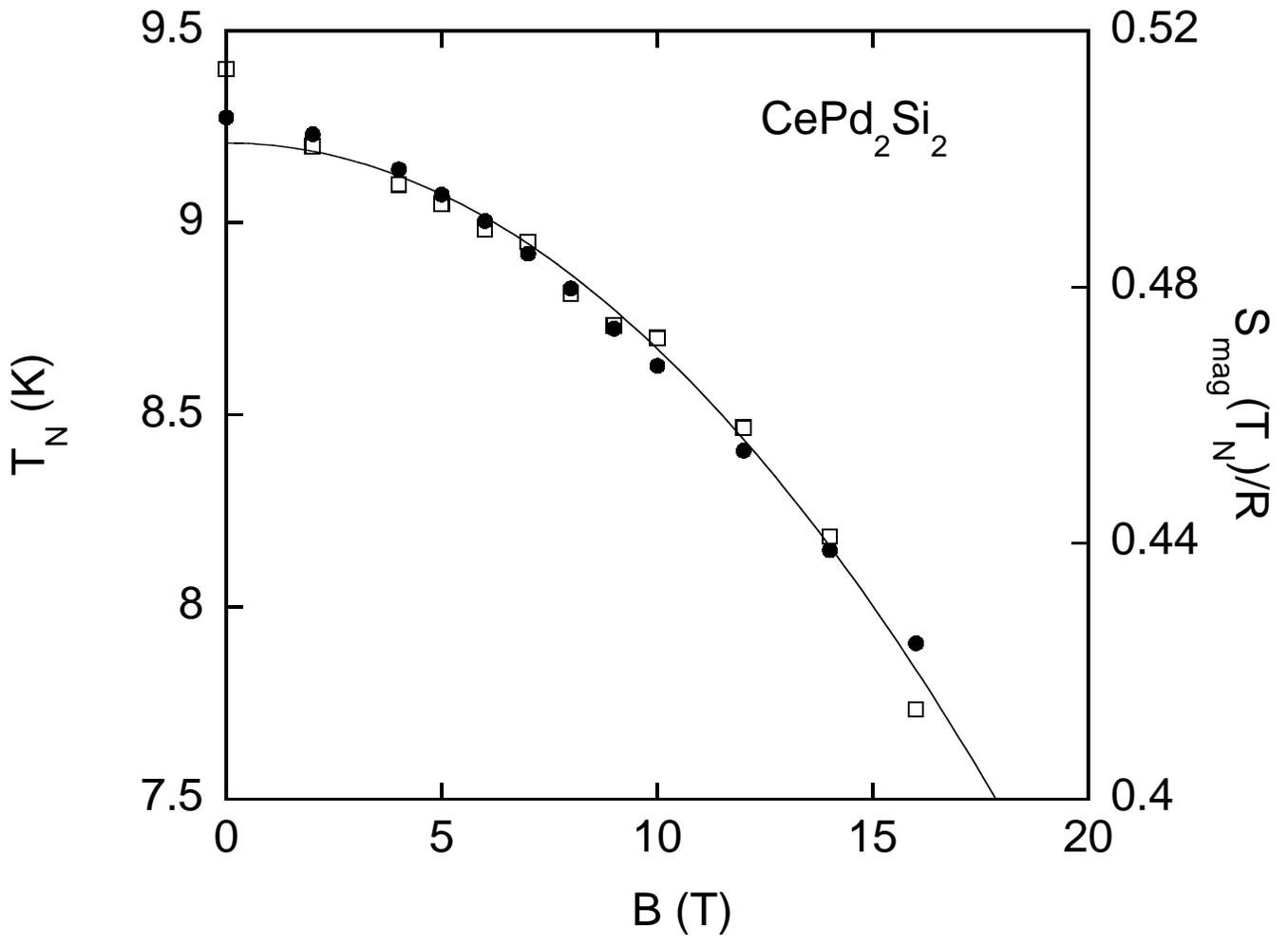
\end{document}